\documentclass[11pt,nofootinbib,floatfix,pra]{revtex4}%
\usepackage{amsfonts,amssymb,stmaryrd,latexsym,amsmath,braket}
\usepackage{graphicx,subfigure}
\usepackage[export]{adjustbox}
\usepackage{times}
\usepackage{bbm}
\usepackage{bm}
\usepackage{mathrsfs}
\usepackage{slashed}
\usepackage[version=3]{mhchem}
\usepackage{braket}
\usepackage{hyperref}
\usepackage{verbatim}
\usepackage[utf8]{inputenc}
\usepackage{tikz}
\usetikzlibrary{patterns}

\usepackage{enumitem}

\usepackage{graphicx,subfigure}
\usepackage{times}
\usepackage{slashed}
\usepackage{braket}
\usepackage{verbatim}
\usepackage{multirow}
\usepackage[utf8]{inputenc}
\usepackage{array}
\usepackage{etoolbox}
\usepackage{mathtools}
\usepackage[percent]{overpic}
\newcommand{\PreserveBackslash}[1]{\let\temp=\\#1\let\\=\temp}
\newcolumntype{C}[1]{>{\PreserveBackslash\centering}p{#1}}
\newcolumntype{R}[1]{>{\PreserveBackslash\raggedleft}p{#1}}
\newcolumntype{L}[1]{>{\PreserveBackslash\raggedright}p{#1}}

\makeatletter
\let\orgdescriptionlabel\descriptionlabel
\renewcommand*{\descriptionlabel}[1]{%
  \let\orglabel\label
  \let\label\@gobble
  \phantomsection
  \edef\@currentlabel{#1}%
  \let\label\orglabel
  \orgdescriptionlabel{#1}%
}
\makeatother

\usepackage{color}

\usepackage{makecell}
\definecolor{DarkMidnightBlue}{rgb}{0.0, 0.04, 0.14}

\def\){\right)} 
\def\({\left(} 
\def\]{\right]} 
\def\[{\left[}

\hypersetup{hidelinks,
breaklinks=true,
colorlinks=true,
linkcolor=blue, 
citecolor=blue, 
urlcolor=DarkMidnightBlue,
bookmarksopen=false,    
pdftitle={Title},
pdfauthor={Author}}

\def\affHISKP{\affiliation{Helmholtz-Institut f\"ur Strahlen- und Kernphysik~(Theorie) and Bethe Center for Theoretical Physics, Universit\"at Bonn, D-53115 Bonn, Germany}}

\def\affKFUPM{\affiliation{King Fahd University of Petroleum and Minerals (KFUPM), 31261 Dhahran, Saudi Arabia}}

\def\affGIBTU{\affiliation{Faculty of Natural Sciences and Engineering, Gaziantep Islam Science and Technology University, Gaziantep 27010, Turkey}}

\def\affJUELICH{\affiliation{
Institute~for~Advanced~Simulation (IAS-4), Forschungszentrum~J\"{u}lich, D-52425~J\"{u}lich,~Germany}}

\def\affPeng{\affiliation{Peng Huanwu Collaborative Center for Research and Education, International Institute for Interdisciplinary and Frontiers, Beihang University, Beijing 100191, China}}

\begin{document}

\title{\textit{Ab initio} lattice study of neutron–alpha scattering with chiral forces at N3LO}

\author{Serdar Elhatisari}
\email{selhatisari@gmail.com}
\affGIBTU\affKFUPM

\author{Fabian~Hildenbrand}
\email{f.hildenbrand@fz-juelich.de}
\affJUELICH

\author{Ulf-G.~Meißner}
\email{meissner@hiskp.uni-bonn.de}
\affHISKP
\affJUELICH
\affPeng

\begin{abstract}
We present the first \emph{ab initio} lattice calculation of neutron--alpha ($n$--$\alpha$) scattering using nuclear lattice effective field theory (NLEFT) with chiral interactions at next-to-next-to-next-to-leading order (N3LO). Building on the high-fidelity chiral Hamiltonian introduced in Ref.~\cite{Elhatisari:2022zrb}, we compute scattering phase shifts in the $S$- and $P$-wave channels using the L{\"u}scher finite-volume method. Our results demonstrate excellent agreement with empirical $R$-matrix phase shifts in the $^2S_{1/2}$ and $^2P_{3/2}$ channels, while revealing persistent discrepancies in the $^2P_{1/2}$ channel for neutron energies above 5~MeV. To systematically investigate these discrepancies, we construct and analyze a simplified neutron--alpha toy model, demonstrating that these discrepancies are not due to the use of the  L{\"u}scher finite-volume method.  
Additionally, we revisit our three-nucleon (3N) force fitting procedure, explicitly incorporating neutron-alpha scattering data through comprehensive Markov Chain Monte Carlo (MCMC) sampling. This analysis confirms the stability of nuclear binding-energy predictions and highlights the need for further refinements in the lattice N3LO three-nucleon forces to fully describe neutron-alpha scattering in the challenging ${}^2P_{1/2}$ channel.

\end{abstract}

\maketitle

\newpage

\section{Introduction
\label{sec:introduction}}

Understanding nucleon–nucleus scattering from first principles is a long-standing research program in nuclear physics, see e.g.~\cite{Quaglioni:2009mn,Navratil:2010jn,Navratil:2011ay,Vorabbi:2025idm}. In particular, neutron–alpha ($n$--$\alpha$) scattering (a neutron scattering off a ${}^4$He nucleus) represents the simplest nucleon–cluster system beyond the deuteron and the triton. This scattering process provides critical benchmarks for nuclear forces and many-body methods~\cite{Lynn:2015jua}. Low-energy $n$--$\alpha$ scattering produces the well-known ${}^5$He resonances (in the $P$-wave channels), and reproducing these resonance energies and widths has been a stringent test for \textit{ab initio} theories~\cite{Quaglioni:2008sm,Kravvaris:2020lhp}. Unlike properties of atomic nuclei (e.g. ground state energies, charge radii or matter distributions), scattering phase shifts probe the long-range tail and the dynamical properties of nuclear interactions. Accurate phase shift predictions are therefore essential to validate nuclear force models beyond just the bulk properties of atomic nuclei and nuclear matter. For example, the splitting between the ${}^2P_{3/2}$ and ${}^2P_{1/2}$ phase shifts in $n$--$\alpha$ scattering is directly sensitive to spin–orbit components of the nuclear force~\cite{Pieper:1993zz} and three-nucleon ($3N$) interactions~\cite{Lynn:2015jua,Hebeler:2020ocj}. Even when two-nucleon forces are constrained to two-nucleon scattering data with a high precision and they give a good description for ${}^4$He nucleus, their predictive power in $n$--$\alpha$ scattering reveals whether the Hamiltonian can simultaneously describe clustering and resonance behavior.

In recent years, substantial progress has been made in \textit{ab initio} calculations of light-nucleus scattering observables. The Faddeev–Yakubovsky formalism has been extended to the five-nucleon system, providing exact benchmarks for $n$--$\alpha$ scattering with phenomenological potentials~\cite{Lazauskas:2017paz,Lazauskas:2019hil,Zhang:2020rhz,Mercenne:2021hef}. Likewise, the no-core shell model with continuum (NCSMC) have been applied to $n$--$\alpha$ elastic scattering, yielding phase shifts and resonance properties from chiral $2N$ and $3N$ forces~\cite{Kravvaris:2020lhp,Shirokov:2018nlj}. For instance, in Ref.~\cite{Kravvaris:2020lhp} calculations quantified the uncertainties in $n$--$\alpha$ phase shifts order-by-order in chiral EFT, demonstrating the impact of uncertainties due to $3N$ forces on the ${}^2P_{1/2}$ and ${}^2P_{3/2}$ splitting. Similarly, in Ref.~\cite{Shirokov:2018nlj} $n$--$\alpha$ scattering using realistic phenomenological interactions (JISP16 and Daejeon16) was studied, obtaining phase shifts and resonance parameters in reasonable agreement with experiment. 
\textit{Ab initio} studies of $n$--$\alpha$ scattering were also performed by using  quantum Monte Carlo techniques~\cite{PhysRevC.36.27,Nollett:2006su,Lynn:2015jua,Yang:2025mhg}. These studies revealed that an accurate description of $n$--$\alpha$  scattering requires including $3N$ forces and treating the continuum explicitly, and including a realistic $3N$ interaction is crucial to accurately reproduce the $P$-wave phase shifts, particularly the spin–orbit splitting between the ${}^5$He resonances. Furthermore, in Ref.~\cite{Yang:2025mhg} it was demonstrated that $3N$ forces are also crucial for the accurate description of peripheral phase shifts in the $D$-wave channel.

From the discussion above, it is well understood that in the state-of-the-art \textit{ab initio} theories $n$--$\alpha$  scattering provides a demanding test which most realistic interactions pass only after $3N$ force effects are incorporated. In particular, while most realistic interactions reproduce the ${}^2P_{3/2}$ resonance (the ground state of ${}^5$He) reasonably well, the resonance in the excited ${}^2P_{1/2}$ channel remains more elusive, with its energy and width highly sensitive to the details of the interactions.

One powerful framework that has emerged for \textit{ab initio} nuclear structure and reactions is Nuclear Lattice Effective Field Theory (NLEFT). In NLEFT, nuclear forces are derived from chiral EFT, and then powerful lattice methods are employed to solve the many-nucleon Schrödinger equation~\cite{Lee:2008fa,Lahde:2019npb}. The Hamiltonian is formulated on a periodic cubic lattice, and observables are computed using stochastic, non-perturbative methods such as auxiliary-field quantum Monte Carlo simulations. Lattice methods differ from continuum \textit{ab initio} methods in that no truncated harmonic oscillator basis is used, instead, space is discretized, and one works with position-space configurations of nucleons. As a result, NLEFT is well suited to describe cluster formation~\cite{Elhatisari:2016owd,Elhatisari:2017eno}, scattering~\cite{Elhatisari:2015iga,Elhatisari:2016hby,Elhatisari:2016hui,Elhatisari:2021eyg}, and fragmentation into subsystems~\cite{Lu:2019nbg,Ren:2023ued}, since in the lattice simulations deformations, polarizations, clustering and disintegration are naturally emerging in the formation of spatially distinct clusters~\cite{Elhatisari:2017eno,Shen:2022bak,Shen:2024qzi}. 

Over the past decade, NLEFT has been successfully applied to a broad range of nuclear problems. Recent progresses have developed a high-fidelity chiral force at N3LO (next-to-next-to-next-to-leading order) that includes the $2N$ interactions up to N3LO~\footnote{Note that in Ref.~\cite{Elhatisari:2022zrb} the two-pion exchange contributions to the NN potential were subsumed in the four-nucleon contact terms.} as well as the  smeared N2LO $3N$ interactions, and a novel method called wavefunction matching~\cite{Elhatisari:2022zrb} to compute observable accurately.  We note that the smearing effectively induces higher order operators
in the $2N$ and the $3N$ sectors, which is of utmost importance.
This action has demonstrated remarkable agreement with experimental data across a wide range of observables, including nucleon–nucleon phase shifts, binding energies and charge radii of nuclei up to $A \leq 58$, the neutron matter equation of state, and the saturation properties of nuclear matter~\cite{Elhatisari:2022zrb}.

It has also been successfully applied to address nuclear systems under diverse physical conditions, structural and thermodynamic properties of hot neutron matter~\cite{Ma:2023ahg}, compute charge radii
of silicon isotopes from ${}^{28}$Si to ${}^{34}$Si~\cite{Konig:2023rwe}, 
calculate the triton lifetime~\cite{Elhatisari:2024otn}, investigate the shell structure and nucleon distributions in proton-rich systems such as ${}^{22}$Si~\cite{Zhang:2024wfd},  study the spectrum and structure of beryllium isotopes from ${}^{7}$Be to ${}^{12}$Be~\cite{Shen:2024qzi}, and study pairing correlations and superfluidity in carbon and oxygen isotopes~\cite{Song:2025ofd}. Furthermore, the NLEFT framework has recently been extended to study the third dimension of the nuclear chart, enabling \textit{ab initio} calculations of hypernuclei and $\Lambda$ separation energies~\cite{Hildenbrand:2024ypw}, using the same N3LO lattice action for the nucleon sector introduced in Ref.~\cite{Elhatisari:2022zrb}.

The wavefunction matching method introduced in Ref.~\cite{Elhatisari:2022zrb} enables rapidly converging first order perturbative calculations using high-fidelity chiral interactions around a non-perturbative Hamiltonian that is almost free of sign oscillations. Along this direction, Ref.~\cite{Lu:2021tab} introduced a perturbative quantum Monte Carlo (ptQMC) method, which allows for second order perturbative calculations for many-body quantum systems. Building on the progress in the NLEFT framework of Ref.~\cite{Li:2018ymw,Elhatisari:2022zrb}, recently a new N3LO lattice force designed for ptQMC simulations has been also formulated~\cite{Liu:2025say,Wu:2025fkn}. Furthermore, the ptQMC framework has been applied to ab initio studies of nuclear beta-decay processes using the chiral interactions at N2LO~\cite{Wang:2025swg}.

The high-fidelity N3LO lattice action introduced in Ref.~\cite{Elhatisari:2022zrb} explicitly includes the subleading $3N$ interactions required by chiral symmetry, such as the two-pion–exchange $3N$ force as well as the one-pion–exchange $3N$ force and the shorter-range $3N$ terms with multiple regulator-defined contact structures known as $c_D$ and $c_E$ interactions. Designing these lattice $3N$ interactions was guided by the framework of cluster effective field theory (cluster EFT), where one treats tightly bound clusters (such as the $\alpha$ particle, i.e. $^4$He) as effective degrees of freedom and builds interactions among them and additional nucleons. This treatment suggested that the lattice action would require multiple independent $3N$ low-energy constants (LECs) to capture the diverse cluster substructures that appear in light nuclei, which was discussed extensively in Ref.~\cite{Elhatisari:2022zrb}. By using the cluster EFT framework, the $c_D$ and $c_E$ terms were refined into several components, each corresponding to a particular smearing and spatial regulator. The strategy was to tune the short-distance parts of the $3N$ force in a minimal but flexible way so that the low-energy observables are free of any regulator artifacts.

The $3N$ forces include the interactions $V_{c_D}^{(d)}$ (for $d = 0, 1, 2$) which denotes locally smeared the one-pion-exchange $3N$ interactions (the $c_D$ term) where $d$ specifies the range of the smearing and the interactions $V_{c_E}^{(d)}$ (for $d = 0, 1, 2$) which are the locally smeared contact $c_E$ interaction. Additionally, there are two additional contact $c_E$ interactions $V_{c_E}^{(l)}$ and $V_{c_E}^{(t)}$ which correspond to different spatial distributions of three nucleons on neighboring lattice sites. By construction, each of these smeared contact interactions shares the same physics (a three-body contact) but with a different regulator profile. Superposing them (with individually fitted coefficients) gives a net $3N$ force that is much less sensitive to any one regulator choice. 

With these smeared and spatially regulated $3N$ interactions, NLEFT is able to effectively encode the additional binding energy or repulsion associated with the geometry. For instance, a shorter-range triangular $3N$ contact can effectively simulate the extra binding of a compact cluster, while a longer-range linear $3N$ term can influence the energy of chain-like configurations. Consistent with this design, in Ref.~\cite{Elhatisari:2022zrb} it was shown that the higher-order $3N$ adjustments indeed play a crucial role in reproducing the properties of atomic nuclei.

A different interpretation of these 3NFs does not require cluster EFT. In fact, one can view the local and non-local smearings as a tool to
include higher derivative operators, as distances on the lattice can be translated into derivatives. Thus, all these smearings corresponds
to terms beyond N2LO in the chiral counting and such terms in general come with new LECs, so that we have more LECs than in the continuum
N2LO theory. Note, however, that a direct mapping on the continuum theory is not easily acheived as in NLEFT one keeps the lattice spacing
finite. To obtain such a dictionary goes beyond the work presented here but would certainly be useful to have.

All eight of the $3N$ forces in Ref.~\cite{Elhatisari:2022zrb} were constrained to ground-state binding energies of selected light and medium-mass nuclei. The guiding principle was that an accurate reproduction of binding energies would indirectly ensure that the $3N$ forces are correctly capturing the net attractive or repulsive effects of three-body correlations in nuclei. The use of binding energies as calibration observables is a pragmatic choice in NLEFT, as these quantities are straightforward to compute with lattice Monte Carlo and they also encapsulate many-body effects. By contrast, experimental few-body scattering data (e.g. nucleon–deuteron scattering or $n$--$\alpha$ scattering observables) are not directly used in the fits, in order to keep the focus on many-body fidelity and to avoid the computational expenses. Going forward, however, an attractive avenue is to further constrain and validate the $3N$ force parameters using scattering information obtained within the lattice framework. 

In this work, we present a complete NLEFT calculation of $n$--$\alpha$ scattering using chiral EFT interactions at N3LO from Ref.~\cite{Elhatisari:2022zrb}. This represents the first ab initio calculations of $n$-–${}^4$He phase shifts on the lattice with an interaction of this precision. By using the L{\"u}scher method, we are able to extract the elastic scattering phase shifts for the relevant partial waves (in particular, the $S$-wave and the ${}^2P_{1/2}$ and ${}^2P_{3/2}$ channels). 
This study not only benchmarks NLEFT’s predictive power in scattering channels but also addresses the applicability of the L{\"u}scher method used to extract the scattering information from finite volumes. In Sect.~\ref{sec:theoretical-framework} we summarize the theoretical framework underlying this investigation.
Sect.~\ref{sec:results-discussions} contains the results and discussion, with a particular emphasis on the applicability of the L\"uscher approach as
well as the role of the spin-orbit splitting in the description of the $n$--$\alpha$ $P$-waves. A summary and outlook is given in Sect.~\ref{sec:summary}.

\section{Theoretical Framework
\label{sec:theoretical-framework}}

In this paper, we use the wavefunction matching method developed in Ref.~\cite{Elhatisari:2022zrb}.
In a nutshell, the full N3LO chiral Hamiltonian $H$ is unitarily transformed so that it can be  matched at some radius to a simple Hamiltonian $H^S$, treating the difference $H-H^S$ in perturbation theory. The simple Hamiltonian is treated non-perturbatively and shows very little sign oscillations, in contrast to the full chiral Hamiltonian. Because of that, one is able to
produce very precise results beyond the accuracy that was possible before.

\subsection{Non-perturbative Hamiltonian
\label{sec:non-pert-Hamiltonian}}

Before presenting our lattice Hamiltonians, we define the important lattice operators used throughout the paper. In our notation $a_{i,j}^{\,}$ and $a_{i,j}^{\dagger}$ are annihilation and creation operators for nucleons with spin index $i=0,1$ (up, down) and isospin index $j=0,1$ (proton, neutron). We also define the non-locally smeared annihilation and creation operators,
\begin{align}
\tilde{a}^{\,}_{i,j}(\vec{n}) = a_{i,j}(\vec{n})+s_{\rm NL}\sum_{|\vec{n}^{\prime}-\vec{n}|=1}a_{i,j}(\vec{n}^{\prime}),
\end{align}
\begin{align}
 \tilde{a}^{\dagger}_{i,j}(\vec{n}) =  a_{i,j}^{\dagger}(\vec{n})+s_{\rm NL}\sum_{|\vec{n}^{\prime}-\vec{n}|=1}a_{i,j}^{\dagger}(\vec{n}^{\prime})\,,
\end{align}
on every lattice site $\vec{n}$  of our lattice with a real parameter $s_{\rm NL}$. Now by defining a real parameter $s_{\rm L}$, we write the most general form of the point-like density operators as, 
\begin{align}
{\rho}^{(d)}(\vec{n}) = & \sum_{i,j} 
{a}^{\dagger}_{i,j}(\vec{n}) \, {a}^{\,}_{i,j}(\vec{n})
+
s_{\rm L}
 \sum_{|\vec{n}-\vec{n}^{\prime}|^2 = 1}^d 
 \,
 \sum_{i,j}  {a}^{\dagger}_{i,j}(\vec{n}^{\prime}) \, {a}^{\,}_{i,j}(\vec{n}^{\prime})
\,,\\
{\rho}^{(d)}_{I}(\vec{n}) =  & \sum_{i,j,j^{\prime}} 
{a}^{\dagger}_{i,j}(\vec{n}) \,\left[\tau_{I}\right]_{j,j^{\prime}} \, {a}^{\,}_{i,j^{\prime}}(\vec{n})
+
s_{\rm L}
 \sum_{|\vec{n}-\vec{n}^{\prime}|^2 = 1}^d 
 \,
  \sum_{i,j,j^{\prime}} 
{a}^{\dagger}_{i,j}(\vec{n}^{\prime}) \,\left[\tau_{I}\right]_{j,j^{\prime}} \, {a}^{\,}_{i,j^{\prime}}(\vec{n}^{\prime})\,, \\
\rho^{(d)}_{S}(\vec{n}) = & \sum_{i,j,i^{\prime}} 
a^{\dagger}_{i,j}(\vec{n}) \, [\sigma_{S}]_{ii^{\prime}}  \, a^{\,}_{i^{\prime},j}(\vec{n}) +
s_{\rm L}
 \sum_{|\vec{n}-\vec{n}^{\prime}|^2 = 1}^d 
 \,
  \sum_{i,j,i^{\prime}} 
a^{\dagger}_{i,j}(\vec{n}^{\prime}) \, [\sigma_{S}]_{ii^{\prime}} \, a^{\,}_{i^{\prime},j}(\vec{n}^{\prime})\,, \\
\rho^{(d)}_{S,I}(\vec{n}) = & \sum_{i,j,i^{\prime},j^{\prime}} 
a^{\dagger}_{i,j}(\vec{n}) \, [\sigma_{S}]_{ii^{\prime}} \, [\sigma_{I}]_{jj^{\prime}} \, a^{\,}_{i^{\prime},j^{\prime}}(\vec{n}) +
s_{\rm L}
 \sum_{|\vec{n}-\vec{n}^{\prime}|^2 = 1}^d 
 \,
  \sum_{i,j,i^{\prime},j^{\prime}} 
a^{\dagger}_{i,j}(\vec{n}^{\prime}) \, [\sigma_{S}]_{ii^{\prime}} \, [\sigma_{I}]_{jj^{\prime}} \, a^{\,}_{i^{\prime},j^{\prime}}(\vec{n}^{\prime})\,,
\end{align}
and the most general form of the non-locally smeared density operators as, 
\begin{align}
\tilde{\rho}^{(d)}(\vec{n}) = & \sum_{i,j} 
\tilde{a}^{\dagger}_{i,j}(\vec{n}) \, \tilde{a}^{\,}_{i,j}(\vec{n})
+
s_{\rm L}
 \sum_{|\vec{n}-\vec{n}^{\prime}|^2 = 1}^d 
 \,
 \sum_{i,j}  \tilde{a}^{\dagger}_{i,j}(\vec{n}^{\prime}) \, \tilde{a}^{\,}_{i,j}(\vec{n}^{\prime})
\,,\\
\tilde{\rho}^{(d)}_{I}(\vec{n}) =  & \sum_{i,j,j^{\prime}} 
\tilde{a}^{\dagger}_{i,j}(\vec{n}) \,\left[\tau_{I}\right]_{j,j^{\prime}} \, \tilde{a}^{\,}_{i,j^{\prime}}(\vec{n})
+
s_{\rm L}
 \sum_{|\vec{n}-\vec{n}^{\prime}|^2 = 1}^d 
 \,
  \sum_{i,j,j^{\prime}} 
\tilde{a}^{\dagger}_{i,j}(\vec{n}^{\prime}) \,\left[\tau_{I}\right]_{j,j^{\prime}} \, \tilde{a}^{\,}_{i,j^{\prime}}(\vec{n}^{\prime})\,, \\
\tilde{\rho}^{(d)}_{S}(\vec{n}) = & \sum_{i,j,i^{\prime}} 
\tilde{a}^{\dagger}_{i,j}(\vec{n}) \, [\sigma_{S}]_{ii^{\prime}}  \, \tilde{a}^{\,}_{i^{\prime},j}(\vec{n}) +
s_{\rm L}
 \sum_{|\vec{n}-\vec{n}^{\prime}|^2 = 1}^d 
 \,
  \sum_{i,j,i^{\prime}} 
\tilde{a}^{\dagger}_{i,j}(\vec{n}^{\prime}) \, [\sigma_{S}]_{ii^{\prime}} \, \tilde{a}^{\,}_{i^{\prime},j}(\vec{n}^{\prime})\,, \\
\tilde{\rho}^{(d)}_{S,I}(\vec{n}) = & \sum_{i,j,i^{\prime},j^{\prime}} 
\tilde{a}^{\dagger}_{i,j}(\vec{n}) \, [\sigma_{S}]_{ii^{\prime}} \, [\sigma_{I}]_{jj^{\prime}} \, \tilde{a}^{\,}_{i^{\prime},j^{\prime}}(\vec{n}) +
s_{\rm L}
 \sum_{|\vec{n}-\vec{n}^{\prime}|^2 = 1}^d 
 \,
  \sum_{i,j,i^{\prime},j^{\prime}} 
\tilde{a}^{\dagger}_{i,j}(\vec{n}^{\prime}) \, [\sigma_{S}]_{ii^{\prime}} \, [\sigma_{I}]_{jj^{\prime}} \, \tilde{a}^{\,}_{i^{\prime},j^{\prime}}(\vec{n}^{\prime})\,,
\end{align}
where the superscript $d$ specifies the range of the local smearing. 

We begin by defining our simplified Hamiltonian, denoted as $H^S$, employed in the non-perturbative part of our lattice calculations. This Hamiltonian is constructed using an approximation to the leading-order chiral EFT interaction, given explicitly by,
\begin{equation}
H^S=K+\frac{1}{2}c_{{\rm SU(4)}}\sum_{\vec{n}}:
\left[
\tilde{\rho}^{(1)}(\vec{n})
\right]^2
:  
+ 
V_{\rm OPE}^{\Lambda_{\pi}} \,,
\label{eq:Hsimple2}
\end{equation}
where $K$ is the kinetic energy operator with the nucleon mass $m_{N}=938.92$~MeV, and $::$ denotes normal ordering.
The kinetic operator utilizes fast Fourier transforms, exactly reproducing the $E_N = p^2/(2m_{N})$ dispersion relation. Throughout our calculations, we employ the smearing parameters $s_{\rm L}=0.07$ and $s_{\rm NL} = 0.5$, values closely following those used previously~\cite{Lu:2018bat,Lu:2019nbg}. Both parameters influence the range of the two-nucleon interactions, but the local smearing $s_{\rm L}$  significantly impacts the alpha-alpha interaction, crucial for nuclear binding~\cite{Elhatisari:2016owd}. Hence, choosing these values carefully ensures that the non-perturbative Hamiltonian $H^S$ captures the essential features of the many-body system under study.

In addition to the short-range SU(4)-symmetric terms, we also incorporate the leading-order long-range one-pion-exchange (OPE) potential, regularized following the procedure from Ref.~\cite{Reinert:2017usi},
\begin{align}
V_{\rm OPE}^{\Lambda_{\pi}}  = - &   \frac{g_A^2}{8F^2_{\pi}}\ \, \sum_{{\vec{n}',\vec{n}},S',S,I}
:\rho_{S',I\rm }^{(0)}(\vec{n}')f_{S',S}(\vec{n}'-\vec{n}) 
  \rho_{S,I}^{(0)}(\vec{n}):  
\,,
\label{eq:OPEP-full}
\end{align}
where $f_{S',S}$ is the locally-regulated pion correlation function with $\vec{q} = \vec{p}- \vec{p\,}^{\prime}$ the momentum transfer ($\vec{p}$ and $\vec{p\,}^{\prime}$ are the relative incoming and outgoing momenta),
\begin{align}
f_{S',S}(\vec{n}'-\vec{n}) 
= &\frac{1}{L^3}\sum_{\vec{q}}
\frac{q_{S'}q_{S} \, e^{-i\vec{q}\cdot(\vec{n}'-\vec{n})-(\vec{q}^2+M^2_{\pi})/{\tilde\Lambda}_{\pi}^2}}{\vec{q}^2 + M_{\pi}^2} \,,
\end{align}
the regulator parameter ${\tilde\Lambda}_{\pi}$,  $g_{A}=1.287$ the axial-vector coupling constant (adjusted to account for the Goldberger-Treiman discrepancy)~\cite{Fettes:1998ud},  $F_{\pi}=92.2$~MeV the pion decay constant and $M_{\pi}=134.98$~MeV the pion mass. It is important to note that the OPE needs to be included in $H^S$ to make the method work, and
further, this is consistent with Weinberg's power counting~\cite{Weinberg:1991um}.

\subsection{N3LO Hamiltonian
\label{sec:non-pert-Hamiltonian}}

Now we provides some details of the high-fidelity Hamiltonian used in our calculations. The lattice Hamiltonian consisting of the $2N$ interactions in the framework of chiral EFT up to N3LO was developed by constraining to match empirical partial wave phase shifts and mixing angles with high accuracy~\cite{Li:2018ymw}. In this work, the $2N$ interactions were constructed by utilizing a non-local smearing parameter $s_{\rm NL}$ defined in configuration space. This approach is also used in the definition of the short-range interaction in the non-perturbative Hamiltonian given in Sec.~\ref{sec:non-pert-Hamiltonian}. Recently in Ref.~\cite{Elhatisari:2022zrb}, the $2N$ interactions were constructed by using a non-local regulator $f_{\Lambda} = \exp[-\sum_{i=1}^{2}({\vec p}_{i}^2 + {\vec p}_{i}^{\,\prime}{}^2)/\Lambda^2]$, where $\vec p_{i}$ and $\vec p_{i}^{\,\prime}$ are the momenta of incoming and outgoing individual nucleons, respectively. In addition, in Ref.~\cite{Elhatisari:2022zrb} the $3N$ interactions were included into the framework, and the LECs of these $3N$ interactions were determined by fitting some selected nuclear binding energies. In this work we use the same high-fidelity Hamiltonian developed in Ref.~\cite{Elhatisari:2022zrb} which is given by, \begin{equation}
H = K + V_{\rm OPE} + V_{\rm C}
+ V_{\rm 3N}^{\rm Q^3}
+ V_{\rm 2N}^{\rm Q^4}
+ W_{\rm 2N}^{\rm Q^4}\,
\label{eq:H-N3LO}
\end{equation}
where $V_{\rm OPE}$ is the one-pion-exchange potential, $V_{\rm C}$ is the Coulomb interaction, $V_{\rm 3N}^{\rm Q^3}$ is the $3N$ potential, $V_{\rm 2N}^{\rm Q^4}$ is the $2N$ short-range interaction at N3LO, and $W_{\rm 2N}^{\rm Q^4}$ is the $2N$ Galilean invariance restoration (GIR) interaction at N3LO. For the details of the Coulomb interaction and the $2N$ short-range interactions we refer the reader to Ref.~\cite{Li:2018ymw}. The kinetic energy operator $K$ is the same as that of the non-perturbative Hamiltonian given in Sec.~\ref{sec:non-pert-Hamiltonian}. The $V_{\rm OPE}$ is the one-pion-exchange potential defined using the regularization method given in Ref.~\cite{Reinert:2017usi},
\begin{align}
V_{\rm OPE}  = - & \frac{g_A^2}{8F^2_{\pi}}\ \, \sum_{{\bf n',n},S',S,I}
:\rho_{S',I\rm }^{(0)}(\vec{n}')f_{S',S}(\vec{n}'-\vec{n})   \rho_{S,I}^{(0)}(\vec{n}):  
\nonumber\\
&- C_{\pi} \, \frac{g_A^2}{8F^2_{\pi}} \sum_{{\bf n',n},S,I}
:\rho_{S,I}^{(0)}(\vec{n}')
f^{\pi}(\vec{n}'-\vec{n})
  \rho_{S,I}^{(0)}(\vec{n}):\,,
\label{eq:OPEP-full}
\end{align}
where $f^{\pi}$ is a local regulator defined in momentum space with $\vec{q} = \vec{p}- \vec{p}^{\prime}$ the momentum transfer ($\vec{p}$ and $\vec{p}^{\prime}$ are the relative incoming and outgoing momenta),
\begin{align}
f^{\pi}(\vec{n}'-\vec{n})
= &
\frac{1}{L^3}
\sum_{\vec{q}}
e^{-i\vec{q}\cdot(\vec{n}'-\vec{n})-(\vec{q}^2+M^2_{\pi})/\Lambda_{\pi}^2}\,.
\end{align}
and $C_{\pi}$ is the coupling constant of the OPE counter term given by,
\begin{align}
C_{\pi} = & -
\frac{\Lambda_{\pi} (\Lambda_{\pi}^2-2M_{\pi}^{2}) + 2\sqrt{\pi} M_{\pi}^3\exp(M_{\pi}^2/\Lambda_{\pi}^{2}){\rm erfc}(M_{\pi}/\Lambda_{\pi})}
{3 \Lambda_{\pi}^3}\,,
\end{align}
with $\Lambda_{\pi}=300$~MeV. The three-nucleon term at ${\rm Q^3}$ consist of the locally smeared contact $c_E$ interaction $V_{c_E}^{(d)}$ (for $d = 0, 1, 2$), the one-pion exchange potential with that the two-nucleon contact term is smeared locally $V_{c_D}^{(d)}$ (for $d = 0, 1, 2$), and the two-pion exchange potential~\cite{Friar:1998zt,Epelbaum:2002vt,Epelbaum:2009zsa}.  Additionally, there are two additional contact $c_E$ interactions $V_{c_E}^{(l)}$ and $V_{c_E}^{(t)}$ which correspond to different spatial distributions of three nucleons on neighboring lattice sites. Therefore, the three-nucleon interactions at ${\rm Q^3}$ has the form
\begin{align}
V_{\rm 3N}^{\rm Q^3}
= 
\sum_{d = 0}^{2} \left[ V_{c_{E}}^{(d)} +
  V_{c_{D}}^{(d)} \right] +   V_{c_{E}}^{(l)} +
  V_{c_{E}}^{(t)} +
+ V_{\rm 3N}^{\rm (TPE)}
\,.
\label{eqn:V_NNLO^3N--001}
\end{align}
Here, we first define the two-pion exchange potential, which can be separated into the following three parts,
\begin{align}
V_{\rm 3N}^{\rm (TPE1)}
= & \frac{c_{3}}{F_{\pi}^{2}}\,
\frac{g_{A}^{2}}{4 F_{\pi}^{2}}
\, \sum_{S,S^{\prime},S^{\prime\prime},I}
\sum_{\vec{n},\vec{n}^{\,\prime},\vec{n}^{\,\prime\prime}}	
\, : \, \rho_{S^{\prime},I}^{(0)}(\vec{n}^{\,\prime}) \,
f_{S^{\prime},S}(\vec{n}^{\,\prime}-\vec{n})
f_{S^{\prime\prime},S}(\vec{n}^{\,\prime\prime}-\vec{n})
\rho_{S^{\prime\prime},I}^{(0)}(\vec{n}^{\,\prime\prime}) \,
\rho^{(0)}(\vec{n})
\, : \,
\label{eqn:V_TPE1^3N--001}
\end{align}
\begin{align}
V_{\rm 3N}^{\rm (TPE2)}
= &
-\frac{2c_{1}}{F_{\pi}^{2}}\,
\frac{g_{A}^{2} \, M_{\pi}^{2}}{4 F_{\pi}^{2}}
\, \sum_{S,S^{\prime},I}
\sum_{\vec{n},\vec{n}^{\,\prime},\vec{n}^{\,\prime\prime}}	
\, : \, \rho_{S^{\prime},I}^{(0)}(\vec{n}^{\,\prime}) \,
f_{S^{\prime}}^{\pi\pi}(\vec{n}^{\,\prime}-\vec{n})
f_{S}^{\pi\pi}(\vec{n}^{\,\prime\prime}-\vec{n})
\rho_{S,I}^{(0)}(\vec{n}^{\,\prime\prime}) \,
\rho^{(0)}(\vec{n})
\, : \, \,,
\label{eqn:V_TPE2^3N--001}
\end{align}
\begin{align}
V_{\rm 3N}^{\rm (TPE3)}
=   \frac{c_{4}}{2F_{\pi}^{2}}
&
\left( \frac{g_{A}}{2 F_{\pi}}\right)^{2}
\sum_{S_{1},S_{2},S_{3}}
\sum_{I_{1},I_{2},I_{3}}
\sum_{S^{\prime},S^{\prime\prime}}
\sum_{\vec{n},\vec{n}^{\,\prime},\vec{n}^{\,\prime\prime}}
\varepsilon_{S_1,S_2,S_3}
\varepsilon_{I_1,I_2,I_3}
\nonumber \\
&
\times	
\, : \, \rho_{S^{\prime},I_{1}}^{(0)}(\vec{n}^{\,\prime}) \,
f_{S^{\prime},S_{1}}(\vec{n}^{\,\prime}-\vec{n})
f_{S^{\prime\prime},S_{2}}(\vec{n}^{\,\prime\prime}-\vec{n})
\rho_{S^{\prime\prime},I_{2}}^{(0)}(\vec{n}^{\,\prime\prime}) \,
\rho_{S_{3},I_{3}}^{(0)}(\vec{n})
\, : \, \,,
\label{eqn:V_TPE3^3N--001}
\end{align}
where the LECs of two-pion exchange potentials are fixed from pion--nucleon scattering data, $c_{1}=-1.10(3)$, $c_{3}=-5.54(6)$ and $c_{4}=4.17(4)$, all in GeV$^{-1}$~\cite{Hoferichter:2015tha}. We now define the one-pion exchange interaction with the two-nucleon contact terms smeared locally,
\begin{align}
V_{c_{D}}^{(d)}
= -\frac{c_{D}^{(d)} \, g_{A}}{4F_{\pi}^{4} \Lambda_{\chi}}\,  \sum_{\vec{n},S,I}
\sum_{\vec{n}^{\,\prime},S^{\prime}}
\, : \,
\rho^{(0)}_{S^{\prime},I}(\vec{n}^{\,\prime})
f_{S^{\prime},S}(\vec{n}^{\,\prime}-\vec{n})
\rho^{(d)}_{S,I}(\vec{n})
\rho^{(d)}(\vec{n})
\, : \,
\,,
\label{eqn:V_cD-sL-001} \end{align}
and the locally smeared contact interactions as, \begin{align}
V_{c_{E}}^{(d)}= \frac{1}{6} c_{E}^{(d)}
\, \sum_{\vec{n},\vec{n}^{\,\prime},\vec{n}^{\,\prime\prime}}
\,
\left[\rho^{(d)}(\vec{n}) \right] ^3\,,
\label{eqn:V_cE-sL-001}      \end{align}
and finally we define two additional SU(4) symmetric potentials denoted by $V_{c_{E}}^{(l)}$ and $V_{c_{E}}^{(t)}$ as,
\begin{align}
V_{c_{E}}^{(l)} = c_{E}^{(l)} \, \sum_{\vec{n},\vec{n}^{\,\prime},\vec{n}^{\,\prime\prime}}
\rho^{(d)}(\vec{n})\,
\rho^{(d)}(\vec{n}^{\,\prime}) \, \rho^{(d)}(\vec{n}^{\,\prime\prime}) \delta_{|\vec{n}-\vec{n}^{\,\prime}|^2,1} \, \,  \delta_{|\vec{n}-\vec{n}^{\,\prime\prime}|^2,1} \, \,  \delta_{|\vec{n}^{\,\prime}-\vec{n}^{\,\prime\prime}|^2,4},  \label{eqn:V_cE-l-001}      \end{align}
\begin{align}
V_{c_{E}}^{(t)} = c_{E}^{(t)}
\, \sum_{\vec{n},\vec{n}^{\,\prime},\vec{n}^{\,\prime\prime}}
\rho^{(d)}(\vec{n})\,
\rho^{(d)}(\vec{n}^{\,\prime}) \, \rho^{(d)}(\vec{n}^{\,\prime\prime}) \delta_{|\vec{n}-\vec{n}^{\,\prime}|^2,2} \, \,  \delta_{|\vec{n}-\vec{n}^{\,\prime\prime}|^2,2} \, \,  \delta_{|\vec{n}^{\,\prime}-\vec{n}^{\,\prime\prime}|^2,2}\,.
\label{eqn:V_cE-t-001}
\end{align}

\subsection{L{\"u}scher's Method
\label{sec:Luescher-Method}}

L{\"u}scher’s finite-volume method~\cite{Luscher1986105,Luscher1991531} is a cornerstone technique for extracting scattering phase shifts from the discrete energy spectra of two-body systems in finite-volumes. In a cubic box with periodic boundary conditions, the allowed momenta of non-interacting two-particle states are quantized. When the particles interact, their energy levels shift relative to these free spectra, and L{\"u}scher’s method provides a quantization condition relating these shifted energy levels to the two-body scattering phase shifts. L{\"u}scher’s finite-volume formalism assumes that the interaction range is finite and significantly smaller than the box size, ensuring that beyond some interaction radius, the two objects propagate as free waves. The original derivation was for two spinless point-like particles in elastic, isotropic $S$-wave scattering, and provided a formula linking energy shifts in the box to the elastic phase shift at that energy. Over the years, this formalism has been systematically generalized to incorporate higher orbital angular momentum, systems with non-zero total momentum (moving frames), particles with intrinsic spin, multichannel scattering cases, the extraction of resonance parameters and properties in a finite volume~\cite{Rummukainen1995397,Bernard0806.4495,PhysRevD.83.114508,PhysRevD.85.014506,Eur.Phys.J.A48.114,PhysRevD.85.114507,PhysRevD.86.094513,PhysRevD.88.034502,PhysRevD.88.094507,PhysRevD.88.114507,PhysRevD.89.074507,Morningstar:2017spu,Bernard1010.6018,PhysRevD.85.014027,Doring1111.0616,PhysRevD.87.014502,Doring2013185}.

Explicitly, L\"{u}scher's relation for two-body scattering phase shifts in a finite periodic cubic box is given by~\cite{Luscher1986105,Luscher1991531},
\begin{equation}
p^{2\ell+1} \cot\delta_{\ell}(p)
=
\begin{dcases}
 \frac{2}{\sqrt{\pi}L}
 S(\eta) & \quad \text{for $\ell = 0$} \,,
\\ 
 \frac{2p^2}{\sqrt{\pi}L}
 S(\eta) & \quad \text{for $\ell = 1$} \,,
\end{dcases} 
\label{eqn:phaseshift-005}
\end{equation}
where $\eta$ is given by
\begin{align}
\eta = \left(\frac{L p}{2\pi}\right)^{2}\,,
\end{align}
and $S(\eta)$ is the regulated three-dimensional zeta function given by,
\begin{align}
S(\eta)=\lim_{\Lambda\rightarrow\infty}\[\sum_{\vec{n}}
\frac{\Theta(\Lambda^2-\vec{n}^2)}{\vec{n}^2-\eta}-4\pi\Lambda
\] .
\label{eqn:Zeta_function-001}
\end{align}
For the $n-\alpha$ system, in the infinite-volume and continuum limit, the relative momentum $p$ is directly related to the two-body energy levels $E_{n-\alpha}^{(\infty)}$ by
\begin{align}
\label{eq:E-2body-infinity}
E^{(\infty)}=\frac{p^2}{2\mu}-B_{{}^4{\rm He}},
\end{align}
where $B_{{}^4{\rm He}}$ denotes the infinite-volume binding energy of the ${}^4{\rm He}$ nucleus, and $\mu$ is the reduced mass of the two-body system. At finite volume and non-zero lattice spacing, Eq.~(\ref{eq:E-2body-infinity}) is modified by discretization and finite-volume artifacts. Due to broken Galilean invariance at finite lattice spacing, the reduced mass of the composite object is not exactly the same as that in continuum limit. These lattice-spacing artifacts can be removed by introducing Galilean restoration interactions or by numerically determining an effective reduced mass that accurately reproduces the correct dispersion relations of the composite objects. Additionally, finite-volume corrections arise from momentum-dependent effects associated with composite objects moving in the periodic box. These finite-volume corrections are accounted for using topological correction factors $\tau(\eta)$, reflecting wave function wrapping effects around the periodic boundaries~\cite{Bour:2011ef}. Incorporating these corrections, Eq.~(\ref{eq:E-2body-infinity}) is generalized to
\begin{align}\label{eq:E12_L}
E_{L}=\frac{p^2}{2\mu}
-B_{L,{}^4{\rm He}} + \tau(\eta)\,
\left[
 B_{L,{}^4{\rm He}}-B_{{}^4{\rm He}}
\right] \,.  
\end{align}
The topological factor $\tau(\eta)$ is given explicitly as
\begin{align}\label{eq:taud}
\tau(\eta) = &
\[{\sum_{\vec{k}}\frac{1}{(\vec{k}^{2}-\eta^{2})^2}}\]^{-1}\sum_{\vec{k}}
\frac{\sum_{i=1}^3\cos(2\pi\alpha\,k_i)}{3(\vec{k}^2-\eta^{2})^{2}} \,,
\end{align}
with $\alpha=1/4$ for the ${}^4{\rm He}$ nucleus comprising four nucleons with equal masses and no further substructure. The topological phase factor given in Eq.~(\ref{eq:E12_L}) was derived for and gives non-negligible improvements for the lowest partial wave~\cite{Bour:2011ef}. For the higher partial waves ($\ell>0$) $\tau(\eta)$ is suppressed by the box size $L$, $\tau(\eta)=\mathcal{O}(L^{-1})$~\cite{Elhatisari:2014lka}, and can be neglected. By numerically solving Eq.~(\ref{eq:E12_L}) self-consistently, one obtains the momentum $p$ for a given finite-volume size $L$ and lattice-determined energies $E_{L}$, $B_{L,{}^4{\rm He}}$ and $B_{{}^4{\rm He}}$.

\section{Results and Discussions\label{sec:results-discussions}}
\subsection{A first look at binding energies and scattering phase shifts}
We use the framework described in Sec.~\ref{sec:theoretical-framework} to compute the ground state energy of the ${}^4$He nucleus and the $\frac{3}{2}{}^-$, $\frac{1}{2}{}^-$, and $\frac{1}{2}{}^+$ states of the $n-\alpha$ (${}^5$He) system using auxiliary-field lattice Monte Carlo simulations. The calculations are performed on a periodic cubic lattice with spatial spacing $a = 1.32$~fm, corresponding to a momentum cutoff $\Lambda = \pi/a \approx 471$~MeV, and temporal lattice spacing $a_{t} = 0.20$~fm. In lattice Monte Carlo simulations for a given box size $L$, the transient energies of low-lying states are calculated as
\begin{equation}\label{eq:transient-energy}
  E_{L,L_{t}} = 
  \frac{\langle \Psi_I| M^{L_{t}/2} \,  H  \, M^{L_{t}/2} | \Psi_I \rangle}{\langle \Psi_I| M^{L_{t}} | \Psi_I \rangle},
\end{equation}
where $\Psi_I$ is the initial wave function, and $M$ is the normal-ordered transfer matrix operator $:e^{- H^{\rm S} a_{t}}:$, with $L_{t}$ the total number of temporal lattice steps. The initial wave functions are single Slater determinants composed of shell-model orbitals. For example, the initial wave function used to calculate the ground state energy of ${}^4$He has two protons and two neutrons occupying the $1s_{1/2}$ orbital. For the $\frac{3}{2}{}^-$ state of ${}^5$He, one extra neutron occupies the $1p_{3/2}$ orbital. Similarly, for the first and second excited states of ${}^5$He, the extra neutron occupies the $1p_{1/2}$ and $2s_{1/2}$ orbitals, respectively.

Lattice Monte Carlo simulations are performed for lattice sizes ranging from $L = 6.60$~fm to $15.8$~fm, with lattice time steps $L_{t}$ ranging from $150$ to $1000$. To extract infinite-time energies, we employ the ansatz given by,
\begin{align}
E_{L,L_{t}}  =   E_{L} +  c_1 e^{-\Delta E_{L} \, \tau} +  c_2 e^{-\Delta E_{L} \, \tau/2},
\label{eqn:Euclidean-time-ext}
\end{align}
where $\tau = Lt \times a_{t}$ is the Euclidean time. The extrapolated energies are presented in Tab.~\ref{tab:energies-in-finite-box} for various lattice sizes.
\begin{table}[htb!]
    \centering
    \begin{tabular}{c|c|c|c|c}
\hline\hline 
 $\quad L~({\rm fm})\quad$      & $\quad E_{L}({}^4{\rm He})~({\rm MeV})\quad$   &  $\quad E_{L}({}^5{\rm He},\frac{1}{2}{}^+)~({\rm MeV}) \quad$ &  $\quad E_{L}({}^5{\rm He},\frac{3}{2}{}^-)~({\rm MeV}) \quad$   & $ \quad E_{L}({}^5{\rm He},\frac{1}{2}{}^-)~({\rm MeV}) \quad$    \\\hline
 $6.60       $   &   $-42.28 (5) $   &   $-26.40 (86) $   &   $-26.41 (109) $   &   $-21.70 (18)$ \\
 $7.92        $   &   $-31.17 (6) $   &   $-26.94 (16) $   &   $-24.79 (51)  $   &   $-20.06 (8)$ \\
 $9.24        $   &   $-29.37 (3) $   &   $-27.92 (13) $   &   $-24.90 (4)   $   &   $-21.90 (10)$ \\
 $10.6        $   &   $-29.39 (3) $   &   $-28.98 (7)  $   &   $-26.08 (3)   $   &   $-23.61 (5)$ \\
 $11.9        $   &   $-29.46 (3) $   &   $-29.43 (14) $   &   $-26.72 (3)   $   &   $-24.41 (7)$ \\
 $13.2       $   &   $-29.49 (2) $   &   $-29.05 (14) $   &   $-27.14 (3)   $   &   $-25.02 (9)$ \\
 $14.5       $   &   $-29.51 (3) $   &   $-29.92 (9)  $   &   $-27.45 (6)   $   &   $-25.36 (10)$ \\
 $15.8       $   &   $-29.54 (3) $   &   $-30.17 (24) $   &   $-27.65 (8)   $   &   $-25.75 (9)$ \\\hline\hline
    \end{tabular}
    \caption{Extrapolated Energies for different states and box sizes employing Eq.~\eqref{eqn:Euclidean-time-ext} for the two-body part of the chiral nuclear force.}
    \label{tab:energies-in-finite-box}
\end{table}
Since the detailed analysis below focuses on the impact of 3N interactions, Tab.~\ref{tab:energies-in-finite-box} presents lattice results obtained using only the 2N chiral interactions at N3LO. The relative momentum of the two-body system, and consequently the scattering phase shifts, are extracted using the L{\"u}scher finite-volume method, see Eq.~(\ref{eqn:phaseshift-005}). The infinite-volume ground state energy of ${}^4$He required for these extractions is obtained following the procedure in Ref.~\cite{Konig:2017krd}. 

\begin{figure}[htb!]
\centering\includegraphics[width=0.49\textwidth]{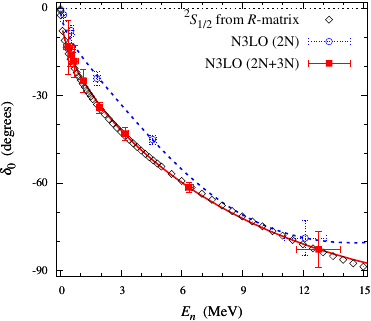}
\hspace{0.5cm}
\centering\includegraphics[width=0.465\textwidth]{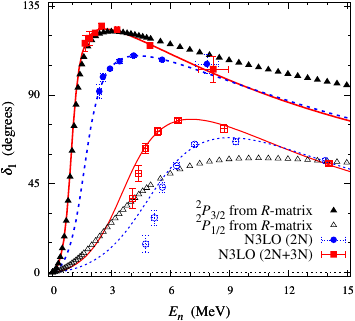}
\caption{
\textbf{Left:}~Neutron-alpha phase shifts in the ${}^2S_{1/2}$ channel computed at N3LO, shown as functions of neutron energy, both with and without the inclusion of 3NFs.
\textbf{Right:}~Neutron-alpha phase shifts in the ${}^2P_{3/2}$ and ${}^2P_{1/2}$ channels computed at N3LO, shown as functions of neutron energy, with and without the inclusion of 3NFs. Empirical phase shifts obtained from an $R$-matrix analysis~\cite{hale2008private} are provided for comparison. Solid colored lines represent fits to the lattice data using the standard effective-range expansion up to $\mathcal{O}(p^6)$.}
\label{fig:n-alpha-Luescher}
\end{figure}
In Fig.~\ref{fig:n-alpha-Luescher}, we show the $n$--$\alpha$ scattering phase shifts computed using only 2N chiral interactions and 2N+3N interactions at N3LO (with 3N forces included up to N2LO and some higher order terms due to the various smearings). For comparison, empirical phase shifts from an $R$-matrix analysis~\cite{hale2008private} are also presented. State-of-the-art N3LO chiral interactions combined with the wavefunction matching method~\cite{Elhatisari:2022zrb} reproduce the empirical $R$-matrix phase shifts remarkably well in the ${}^2P_{3/2}$ and ${}^2S_{1/2}$ channels. However, a clear discrepancy is observed in the ${}^2P_{1/2}$ channel. Interestingly, the ${}^2P_{1/2}$ channel scattering phase shifts exhibit a shape similar to the ${}^2P_{3/2}$ channel but with notably reduced attraction. Additionally, the spin-orbit splitting between the ${}^2P_{3/2}$ and ${}^2P_{1/2}$ channels is already enhanced significantly by 2N chiral interactions alone.

The good agreement observed in the ${}^2S_{1/2}$ channel can be attributed primarily to Pauli repulsion, which is relatively insensitive to the detailed interaction structure. Thus, even 2N chiral interactions alone provide a satisfactory description, and the inclusion of repulsive 3N forces further improves agreement with the data.

In contrast, the good agreement in the ${}^2P_{3/2}$ channel is more profound and arises directly from the fitting strategy employed in Ref.~\cite{Elhatisari:2022zrb} for determining the 3N LECs. Guided by cluster EFT, this strategy treats tightly bound clusters (such as the $\alpha$ particle) as effective degrees of freedom, building interactions among these clusters and additional nucleons. This perspective suggests that different 3N LECs are required to accurately capture the diverse cluster substructures observed in nuclei like ${}^6$He, ${}^9$Be, and ${}^{10}$Be, where extra neutrons predominantly occupy the $1p_{3/2}$ orbital. Consequently, including nuclei such as ${}^6$He, ${}^9$Be, and ${}^{10}$Be in the fitting procedure implicitly propagates effects into scattering observables in the ${}^2P_{3/2}$ channel.  
Note further that the diminishing influence of the 3NFs in the ${}^2P_{3/2}$ and ${}^2P_{1/2}$ channels above
approximately 7~MeV and 9~MeV, respectively, can be understood from the fact that the scattering
physics at higher energies becomes dominated by short-range 2N interactions, causing the
effects of the 3NFs to become relatively suppressed. Additionally, since our lattice 3N
interactions were constructed primarily to reproduce nuclear binding energies, the observed
suppression at higher energies is consistent with this design.

Despite the successful description of the ${}^2P_{3/2}$ channel, the persistent discrepancies observed in the ${}^2P_{1/2}$ channel raise important questions about the origin of these deviations. Specifically, it remains unclear whether these discrepancies stem from the interaction or from the limitations of the phase shift extraction method itself, the L{\"u}scher finite-volume method. To systematically pin down the problem, we perform a detailed study of the L{\"u}scher approach by constructing and analyzing a carefully designed simplified two-body toy model.

\subsection{Neutron--Alpha Toy Model}

To systematically address the observed discrepancy in the ${}^2P_{1/2}$ scattering channel, we construct a simplified neutron--alpha toy model. This model allows us to explicitly examine potential limitations of  L{\"u}scher's finite-volume method when the energy above threshold becomes large.

We consider a system consisting of a spin-$\frac{1}{2}$ neutron and a point-like, spinless alpha particle on a periodic cubic lattice of length $L$. The interaction for this system is defined as,
\begin{equation}
V = \mathbb{I}_{2\times2} \otimes \left[ g_{\rm 0} \,  \delta(r) + g_{\rm c} \exp\left(-\frac{r^2}{R_{\rm c}^2}\right) \right]+ g_{\rm so} \exp\left[-\frac{r^2}{R_{\rm so}^2}\right] \vec{\ell}\cdot\vec{s},
\label{eqn:full-potential}
\end{equation}
where we have a zero-range interaction with an interaction strength $g_{0}$ and restricted to $S$-wave only, a Gaussian central potential with a strength $g_{\rm c}$ and range $R_{\rm c}$, and a spin-orbit interaction with a strength $g_{\rm so}$ and a Gaussian form factor parameterized by $R_{\rm so}$. Also, $\vec{\ell}$ is the orbital angular momentum operator, and $\vec{s}$ is the spin operator. In these calculations we also set the lattice spacing to $a = 1.97$~fm. To determine the interaction parameters $g_{0}$, $g_{\rm c}$, $g_{\rm so}$, $R_{\rm c}$ and $R_{\rm so}$, we compute the $n-\alpha$ phase shifts on the lattice using the 
spherical wall method~\cite{Carlson:1984zz,Borasoy:2007vy} and perform a chi-square fit to the empirical phase shifts from the $R$-matrix analysis~\cite{hale2008private}.

\begin{figure}[htb!]
\centering\includegraphics[width=0.47\textwidth]{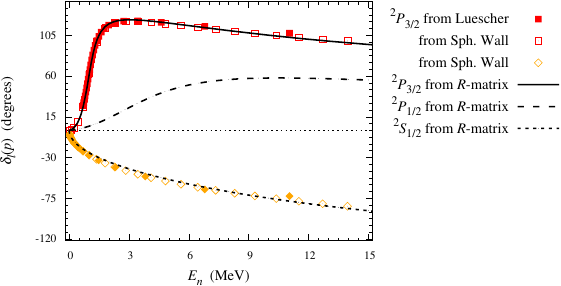}
\hspace{0.5cm}
\centering\includegraphics[width=0.47\textwidth]{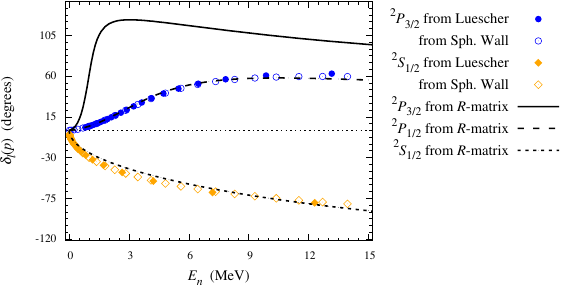}
\caption{\textbf{Left:}~Phase shifts calculated in the ${}^2P_{3/2}$ and ${}^2S_{1/2}$ channels for the neutron-alpha toy model as functions of relative momenta, using the L{\"u}scher (filled symbols) and spherical wall (open symbols) methods.
\textbf{Right:}~Phase shifts calculated in the ${}^2P_{1/2}$ and ${}^2S_{1/2}$ channels for the neutron-alpha toy model as functions of relative momenta, using the L{\"u}scher (filled symbols) and spherical wall (open symbols) methods. Empirical phase shifts from an $R$-matrix analysis~\cite{hale2008private} are also included for comparison.}
\label{fig:n-alpha-toy-model-no-SO-coupling}
\end{figure}

In the first step, we remove the spin-orbit term from the potential given in Eq.~(\ref{eqn:full-potential}) and perform lattice calculations to validate the results from the L{\"u}scher method by benchmarking it with the results from the spherical wall method.
\begin{table}[htb!]
    \centering
    \begin{tabular}{l|c|c|c|c|c}
\hline\hline 
 Channels      & $g_{0}$   & $g_{c}$   &  $R_{\rm c}$      & $g_{\rm so}$   &  $R_{\rm so}$     \\\hline
${}^2S_{1/2}$, ${}^2P_{3/2}$   &   $~~~3.3786\times10^{-1}$ & $-3.4227\times10^{-1}$  & $1.7825$  &  --   & -- \\\hline
${}^2S_{1/2}$, \,${}^2P_{1/2}$ &  $-1.6276\times10^{-1}$ & $-2.0485\times10^{-1}$  & $1.8388$  &  --   & -- \\\hline
${}^2S_{1/2}$, \,${}^2P_{3/2}$, \,${}^2P_{1/2}$  & $~~~9.3623\times10^{-2}$ & $-2.8529\times10^{-1}$  & $1.8254$   &   $-1.0000\times10^{-1}$   & $1.6497$ \\\hline\hline
    \end{tabular}
    \caption{The determined interaction parameters $g_{0}$, $g_{\rm c}$, $g_{\rm so}$, $R_{\rm c}$ and $R_{\rm so}$  for the $n$--$\alpha$ phase shifts on the lattice using the spherical wall method and fitted to the empirical phase shifts from the $R$-matrix analysis. All parameters are given in lattice units.}
    \label{tab:parameters-toy-model}
\end{table}
The interaction parameters determined by fitting the lattice phase shifts obtained using the spherical wall method to the empirical $R$-matrix phase shifts are listed in Table~\ref{tab:parameters-toy-model}. The corresponding calculated phase shifts from both the spherical wall method and the L{\"u}scher method are shown in Figure~\ref{fig:n-alpha-toy-model-no-SO-coupling}. As observed, the phase shifts extracted using the two methods match exactly, confirming the fundamental consistency of the L{\"u}scher approach in the absence of spin-orbit effects.

Next, we turn on the spin-orbit interaction and tune its strength to precisely reproduce the empirical phase shifts in the ${}^2S_{1/2}$, ${}^2P_{3/2}$, and ${}^2P_{1/2}$ channels using spherical-wall calculations. Applying the L{\"u}scher finite-volume method to this case, we find that it accurately reproduces the spherical-wall results in all channels for the energies considered. 
\begin{figure}[htb!]
\centering\includegraphics[width=0.8\textwidth]{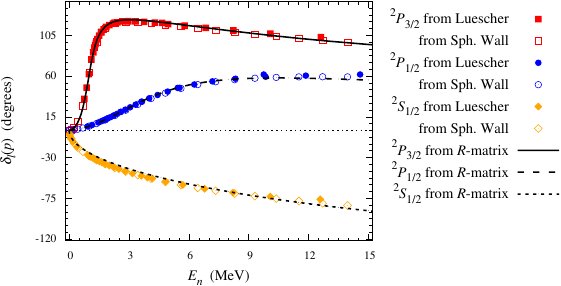}
\caption{Phase shifts calculated for the neutron-alpha toy model with the spin-orbit interaction included, plotted as functions of relative momenta in the ${}^2S_{1/2}$, ${}^2P_{3/2}$, and ${}^2P_{1/2}$ channels. Results obtained from the L{\"u}scher finite-volume method (filled symbols) and spherical wall method (open symbols) are shown, together with empirical phase shifts from an $R$-matrix analysis~\cite{hale2008private} for comparison.}
\label{fig:n-alpha-toy-model}
\end{figure}
So we conclude that the L\"uscher method works fine for the range of neutron energies considered here.

These findings demonstrate that the observed discrepancies in the ${}^2P_{1/2}$ channel in the NLEFT calculation are not due to limitations of the L{\"u}scher finite-volume approach. Thus, to understand the reason behind the discrepancies in the ${}^2P_{1/2}$ channel we also carefully revisit our fitting procedure for the 3NFs. By explicitly incorporating neutron-alpha scattering phase shifts into our fitting procedure, we can examine whether the overall results are sensitive to the specific choice of 3NF parameterizations. We will continue extracting phase shifts using the L{\"u}scher method as it works well here, so this approach will be useful for identifying any potential contributions from the interaction itself to the observed discrepancies in the ${}^2P_{1/2}$ channel.

\subsection{Revisiting the 3NF parameterizations including scattering data}

In this subsection, we revisit the determination of the LECs of the 3NFs originally performed in Ref.~\cite{Elhatisari:2022zrb}. Motivated by our previous findings of the applicability of the L{\"u}scher finite-volume method, we explicitly include neutron-alpha scattering phase shifts into the fitting procedure, alongside the previously considered nuclear binding energies. This approach allows us to directly probe the sensitivity of the neutron-alpha scattering observables, particularly the ${}^2P_{1/2}$ resonance width, to variations of the 3NF parameters.

To ensure a comprehensive and unbiased analysis, we systematically explore subsets of the available eight 3NF terms defined in Eq.~(\ref{eqn:V_NNLO^3N--001}). Specifically, we consider fitting procedures involving combinations of six, seven, and all eight 3NF terms. For each combination, we use Markov chain Monte Carlo (MCMC) sampling, simultaneously optimizing selected observables and subsets of the 3NF interaction parameters. Technical details of this MCMC procedure are beyond the scope of the current discussion, and the method will be presented in a forthcoming publication~\cite{Elhatisari:inprogress}.

Initially, we conduct MCMC sampling with all eight 3NF terms, and the optimization varies only the choice of observables. Subsequently, we perform similar analyses using subsets of seven and six 3NF terms, also varying observable selections. We systematically evaluate the outcomes based on the root mean square deviations (RMSD) relative to experimental data. This approach allowed us to rigorously quantify uncertainties due to both observable selection and specific choices of 3NF parameterizations. For each 3NF combination (six, seven, and eight terms), we retained the top 150 fits with the lowest RMSD values to rigorously quantify systematic uncertainties.

\begin{figure}[htb!]
    \centering
    \begin{overpic}[width=1.0\textwidth]{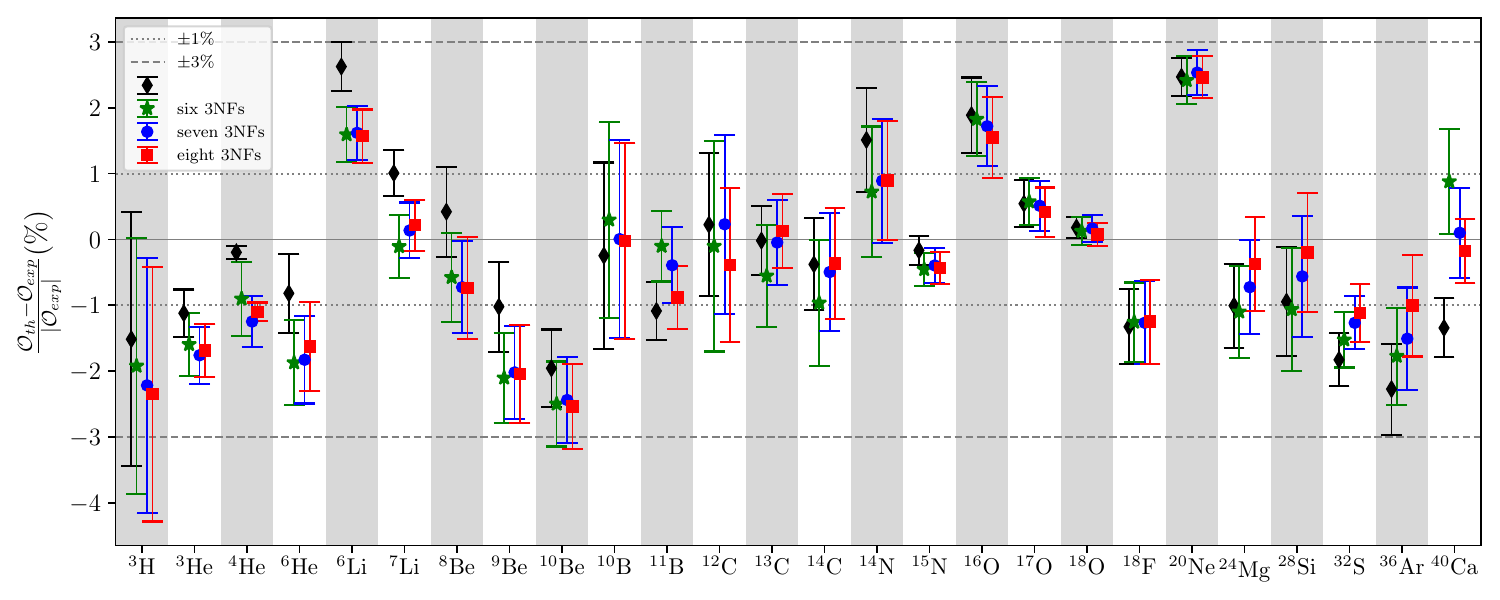}
        \put(11.7,34.0){\tiny Ref.~\cite{Elhatisari:2022zrb}}
    \end{overpic}
    \caption{Plots of the relative deviations between calculated binding energies and experimental data. To benchmark the results of this work we also show the results from Ref.~\cite{Elhatisari:2022zrb}.}
    \label{fig:relative-deviation-BE}
\end{figure}
The results of this refined fitting procedure are presented in Fig.~\ref{fig:relative-deviation-BE}, showing relative deviations (in percent) between calculated nuclear binding energies and experimental values. For validation, results from Ref.~\cite{Elhatisari:2022zrb}, depicted as black diamonds, include statistical uncertainties from lattice simulations, Euclidean time extrapolations, and systematic uncertainties from chiral interactions estimated through history matching analysis.

Our new results, shown by green stars (six 3NF terms), blue circles (seven 3NF terms), and red squares (eight 3NF terms), incorporate statistical uncertainties from lattice simulations and Euclidean time extrapolations, along with systematic uncertainties from observable selections and variations of 3NF parameters. Notably, the outcomes obtained with the full set of eight 3NF terms closely match previous results~\cite{Elhatisari:2022zrb}, with slightly increased uncertainties reflecting the explicit consideration of additional systematic effects. Importantly, results from all tested configurations fall within a narrow 1-2\% error band compared to experimental values. The results presented in Fig.~\ref{fig:relative-deviation-BE} thus validate the extended approach used in this work, demonstrating an accurate reproduction of nuclear binding energies under variations of the 3NF parameterization.

\begin{figure}[htb!]
\centering\includegraphics[width=0.55\textwidth]{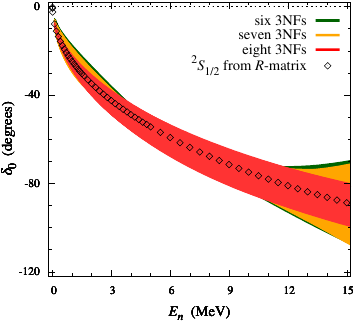}
\caption{Neutron-alpha phase shifts in the ${}^2S_{1/2}$ channel computed at N3LO, shown as functions of neutron energy. The colored band is the smoothed lattice data by fitting to the standard effective-range expansion up to $\mathcal{O}(p^6)$ and they  represent the range of variations in phase shifts obtained from the top 150 parameter sets with the lowest RMSD values. Empirical phase shifts from an $R$-matrix analysis~\cite{hale2008private} are included for comparison.}
\label{fig:n-alpha-refined_S}
\end{figure}

\begin{figure}[htb!]
\centering\includegraphics[width=0.55\textwidth]{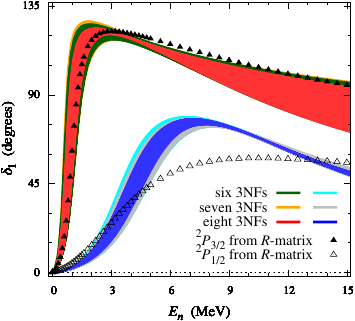}
\caption{Neutron-alpha phase shifts in the ${}^2P_{3/2}$ and ${}^2P_{1/2}$ channels computed at N3LO, shown as functions of neutron energy. The colored bands are the smoothed lattice data by fitting to the standard effective-range expansion up to $\mathcal{O}(p^6)$ and they  represent the range of variations in phase shifts obtained from the top 150 parameter sets with the lowest RMSD values. Empirical phase shifts from an $R$-matrix analysis~\cite{hale2008private} are included for comparison.}
\label{fig:n-alpha-refined}
\end{figure}

We now turn our attention to the results of the refined fitting procedure for the neutron-alpha phase shifts, which explicitly incorporates neutron-alpha scattering data, as discussed previously. The final results, including all systematic and statistical uncertainties, are shown in Figs.~\ref{fig:n-alpha-refined_S} and~\ref{fig:n-alpha-refined}. 
The colored bands illustrate the total variation arising from our MCMC sampling, representing uncertainties associated with both the choice of observables and the selected sets of 3N force terms in the fitting procedure.

Not surprisingly, the earlier seen agreement in the ${}^2S_{1/2}$ channel is validated, only with a better estimate of the uncertainties.
The uncertainties are generally small, with the expected trend that they increase with increasing neutron energy. Also, the more 3NFs are included, the smaller the uncertainty.

Now, let us turn to the $P$-waves. An immediate and notable observation from Fig.~\ref{fig:n-alpha-refined} is that the peak positions in both the ${}^2P_{3/2}$ and ${}^2P_{1/2}$ channels consistently move either upward or downward simultaneously, as the sets of fitted observables and 3N force components are varied. Furthermore, despite the fact that we consider parameters from 150 different fits with the lowest RMSD values, variations in the extracted phase shifts remain relatively small, confirming consistency among different parameterizations. The  discrepancy in the ${}^2P_{1/2}$ channel observed around neutron energies starting at approximately 5~MeV is persistent for all variations of the 3NFs considered here.
This remarkable consistency in the shape and resonance structure of the ${}^2P_{1/2}$ phase shifts across different parameterizations indicates clearly that the fitting procedure itself is not the primary source of the issue. Rather, the form of the lattice three-nucleon interactions currently employed, guided by the principles of cluster effective field theory in Ref.~\cite{Elhatisari:2022zrb}, appears insufficiently flexible to accurately capture the detailed dynamics of the neutron-alpha ${}^2P_{1/2}$ channel, even when explicitly including scattering data in the fits. Consequently, future studies  should focus on improving the form and parameterization of lattice three-nucleon interactions at N3LO to reliably reproduce neutron-alpha scattering observables, particularly in the challenging ${}^2P_{1/2}$ channel.

\section{Summary\label{sec:summary}}

In this study, we presented comprehensive lattice Monte Carlo simulations of neutron-alpha scattering using high-fidelity chiral interactions at N3LO, explicitly incorporating three-nucleon forces (3NFs) with smeared N2LO operator structure, thus including a number of higher order 3NFs. Using a cubic periodic lattice framework, we calculated the ground-state energy of the ${}^4$He nucleus and the relevant low-lying states of the neutron-alpha (${}^5$He) system. We extracted neutron-alpha phase shifts for the ${}^2S_{1/2}$, ${}^2P_{3/2}$, and ${}^2P_{1/2}$ channels via  L{\"u}scher's finite-volume method, systematically exploring the accuracy and limitations of this method.

Our results demonstrated excellent agreement with empirical data for the ${}^2S_{1/2}$ and ${}^2P_{3/2}$ channels, validating both the robustness of our lattice simulations and the predictive power of the employed N3LO chiral interactions. However, a visible discrepancy emerged in the ${}^2P_{1/2}$ channel, particularly at neutron energies above approximately 5~MeV. To systematically investigate this issue, we constructed a simplified neutron-alpha toy model, enabling a direct assessment of the L{\"u}scher method. This analysis clearly revealed that 
the L{\"u}scher finite-volume approach has no inherent methodological limitations for the energies
considered here.

Motivated by these findings, we revisited the fitting procedure for determining the 3NF parameters, explicitly incorporating neutron-alpha scattering data into our fits. Our refined fitting approach, utilizing comprehensive Markov chain Monte Carlo sampling over subsets of the 3NF terms, confirmed the robustness of nuclear binding-energy predictions, showing minimal sensitivity to parameter variations. 
First, the description of the ${}^2S_{1/2}$ stays excellent with small uncertainties. Second,
while the inclusion of scattering data slightly shifted the resonance positions simultaneously in both ${}^2P_{3/2}$ and ${}^2P_{1/2}$ channels, the persistent discrepancy in the ${}^2P_{1/2}$ resonance was unchanged, reaffirming our conclusions about limitations in the present representation of the 3NFs. 

Given that the employed N3LO interactions accurately describe other channels, the unresolved issue in the ${}^2P_{1/2}$ channel underscores the critical need for improvements in the 3NFs. Thus, corresponding  studies will be essential to conclusively assess the accuracy and predictive capabilities of our lattice chiral interactions, ultimately advancing our understanding of nuclear interactions and scattering processes.

\acknowledgements
We are grateful for discussions with members of the Nuclear Lattice Effective Field Theory Collaboration. SE particularly acknowledges helpful discussions with Dean Lee on the Lüscher finite-volume method. We acknowledge funding in part by the Scientific and Technological Research Council of Turkey (TUBITAK
project no. 120F341) and by the European Research Council (ERC) under the European Union's Horizon 2020 research and innovation programme (grant agreement No. 101018170).
The work of UGM was supported in part by the CAS President's International Fellowship Initiative (PIFI) (Grant No.~2025PD0022). Computational resources  provided by TUBITAK ULAKBIM High Performance and Grid Computing Center (TRUBA resources) as well as the Gauss Centre for Supercomputing e.V. (www.gauss-centre.eu) for computing time on the GCS Supercomputer JUWELS at J{\"u}lich Supercomputing Centre (JSC) are gratefully
acknowledged. The authors also gratefully acknowledge the computing time provided on the high-performance
computer HoreKa by the National High-Performance Computing Center at KIT (NHR@KIT). This
center is jointly supported by the Federal Ministry of Education and Research and the
Ministry of Science, Research and the Arts of Baden-Württemberg, as part of the National
High-Performance Computing (NHR) joint funding program
(https://www.nhr-verein.de/en/our-partners). HoreKa is partly funded by the German
Research Foundation (DFG).

\bibliography{References}
\bibliographystyle{apsrev}

\end{document}